%% file: calm-bp-arxiv.tex
\documentclass[11pt]{article}

\usepackage{amsmath}
\usepackage{amssymb}
\usepackage{graphicx}
\usepackage{booktabs}
\usepackage{multirow}
\usepackage[table]{xcolor}
\usepackage{pifont}
\usepackage[margin=1in]{geometry}
\usepackage[hidelinks]{hyperref}

\title{CALM-BP: Observation-Matched Physiological Semantic Grounding for Non-Contact Blood Pressure Estimation}

\author{Sun Haiyang\textsuperscript{1}, Gu Boyuan\textsuperscript{1}, Liu Yongjie\textsuperscript{1}}

\date{}

\begin{document}

\maketitle

\noindent\textsuperscript{1}University of Electronic Science and Technology of China, Chengdu, China.

\begin{abstract}
Language grounding increasingly involves non-text observations whose structure is not naturally expressed as words or objects. We study this problem for physiological time series in non-contact blood pressure (BP) estimation: remote photoplethysmography (rPPG) provides measured evidence about bodily state, but numerical pipelines expose little semantic structure about why a window is reliable or how its cues should be fused. We introduce observation-matched physiological semantic grounding, where language-derived priors must be constructed from the same rPPG observation, remain bounded by an auditable prior contract, and avoid BP-label or identity leakage. CALM-BP does not treat language as new physiological evidence; instead, it verbalizes rPPG descriptors into a controlled semantic interface while rPPG remains the primary haemodynamic evidence source. FlowBP-Set pairs forehead observations, synchronized BP labels, and structured physiological prompts from 81 participants. Main BP results, direct cross-dataset evaluation, language-realization ablation, and observation-mismatch controls test whether language helps because it organizes the current physiological observation rather than because it is arbitrary auxiliary text. The FlowBP-Set dataset contains sensitive facial video and physiological recordings and is therefore not publicly available due to privacy and ethical restrictions. Data access may be considered upon reasonable request and subject to applicable ethical and institutional approval.

\end{abstract}


\section*{Highlights}
\begin{itemize}
\item A bounded language prior verbalizes the same rPPG window as fusion guidance
\item CALM-BP keeps rPPG as primary evidence for non-contact blood pressure estimation
\item FlowBP-Set pairs forehead observations with synchronized BP labels and prompts
\item Grounding controls show gains come from matched priors, not leakage or text
\end{itemize}


\section{Introduction}
\label{sec:intro}

BP is an important physiological indicator for cardiovascular assessment, yet conventional cuff-based or contact measurements are intermittent, inconvenient, and difficult to use for frequent unobtrusive observation. Non-contact BP estimation from rPPG offers a complementary route by extracting pulse-related traces from facial video. This setting is also a useful language-grounding problem: facial appearance is not BP evidence by itself, and inference must depend on haemodynamic traces whose quality varies with region selection, motion, illumination, temporal consistency, and recording state \cite{Poh2010RemotePulse,Wang2017AlgRemotePPG}.

\begin{figure}[t]
    \centering
    \includegraphics[width=\columnwidth]{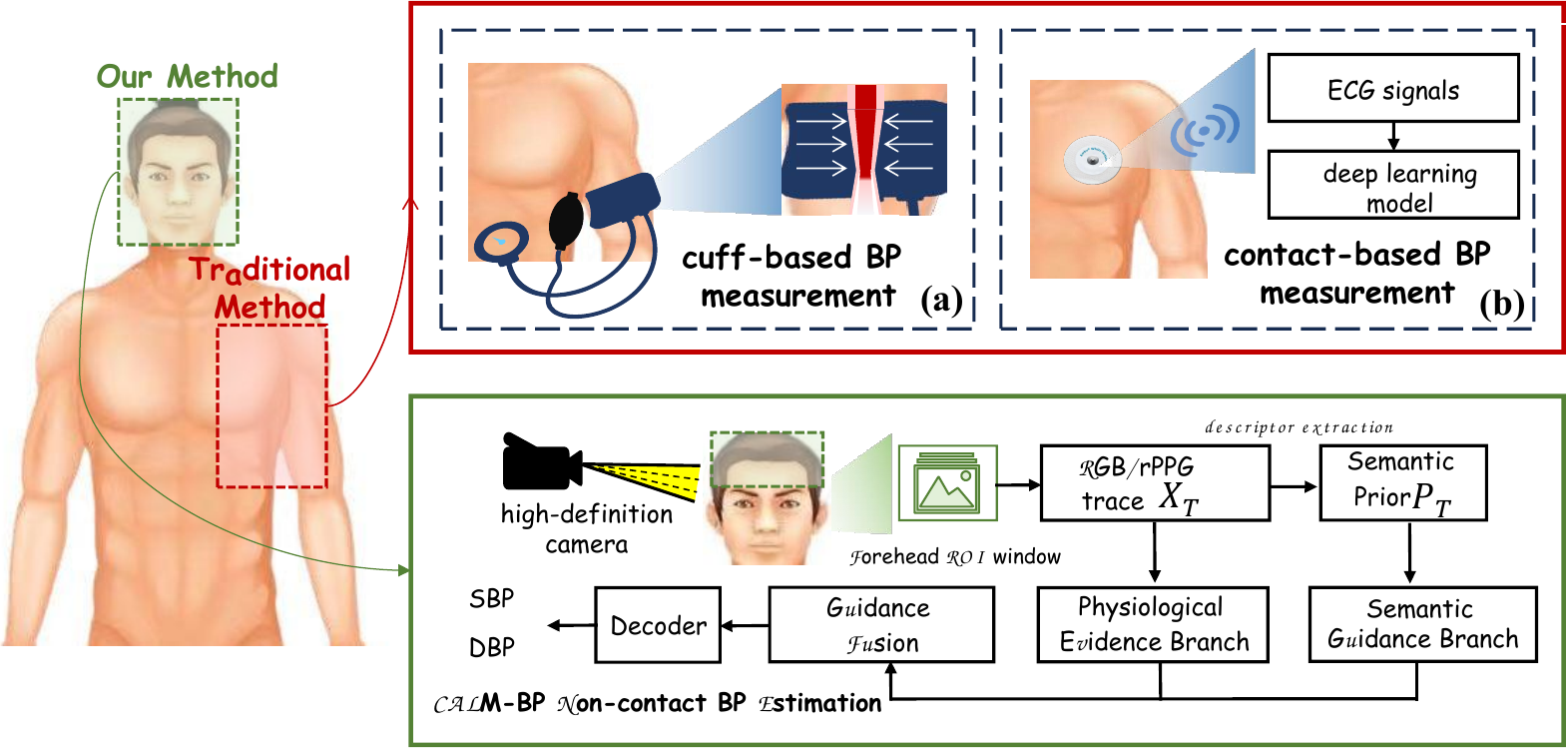}
    \caption{\textbf{FlowBP-Set reference and observation setting.}
    The dataset pairs synchronized reference BP acquisition with non-contact forehead video observations.
    CALM-BP converts the forehead observation into an RGB/rPPG trace $X_T$ as physiological evidence and derives a semantic prior $P_T$ as fusion guidance.
    Neither facial appearance nor language alone is used to infer BP.}
    \label{fig:flowbp_setting}
\end{figure}

Traditional rPPG-based BP pipelines have improved the numerical prediction path through visual, rPPG, and physiological signal modeling \cite{Chen2023RemoteBP,Wu2022rPPGBP,Chen2024DLrPPGReview}. Their limitation is that the semantic context of a window is usually left implicit. Without such context, a model has little explicit structure for separating measured signal evidence from quality, reliability, waveform, and state-aware cues; this makes the prediction path harder to audit and makes it difficult to test whether a model relies on the current physiological observation or on coarse shortcuts. This finding motivates a bounded language interface over non-text physiological observations, consistent with recent work on language-aligned time series and sensor observations \cite{Liu2024TimeCMA,Li2025SensorLLM,Niu2025ProMedTS}.

We establish this setting with FlowBP-Set, a paired resource of forehead observations, synchronized BP references, and structured physiological prompts. Figure~\ref{fig:flowbp_setting} illustrates the reference-and-observation setup. The dataset lets us study what happens when a semantic prior is attached to the same physiological window, detached from it, or evaluated under subject-disjoint splits. This design turns the missing-context problem into a testable grounding setting: a language prior is useful only if it organizes the current measured observation rather than leaking identity, target labels, or coarse protocol shortcuts.

Based on this semantic formulation, we introduce CALM-BP, a \underline{c}ross-\underline{al}igned \underline{m}odel for \underline{b}lood \underline{p}ressure estimation. Rather than asking a multimodal large language model (MLLM) to infer BP directly from appearance, CALM-BP extracts numerical rPPG features from forehead observations and builds promptized physiological priors under an explicit semantic prior contract. A frozen large language model (LLM) encodes these priors, and cross-modal alignment lets the physiological branch retrieve observation-matched semantics before SBP and DBP inference. The physiological pathway remains the evidence source; language contributes a structured, testable fusion interface over signal variation, stability, waveform shape, and state.

\begin{figure*}[t]
    \centering
    \includegraphics[width=\textwidth]{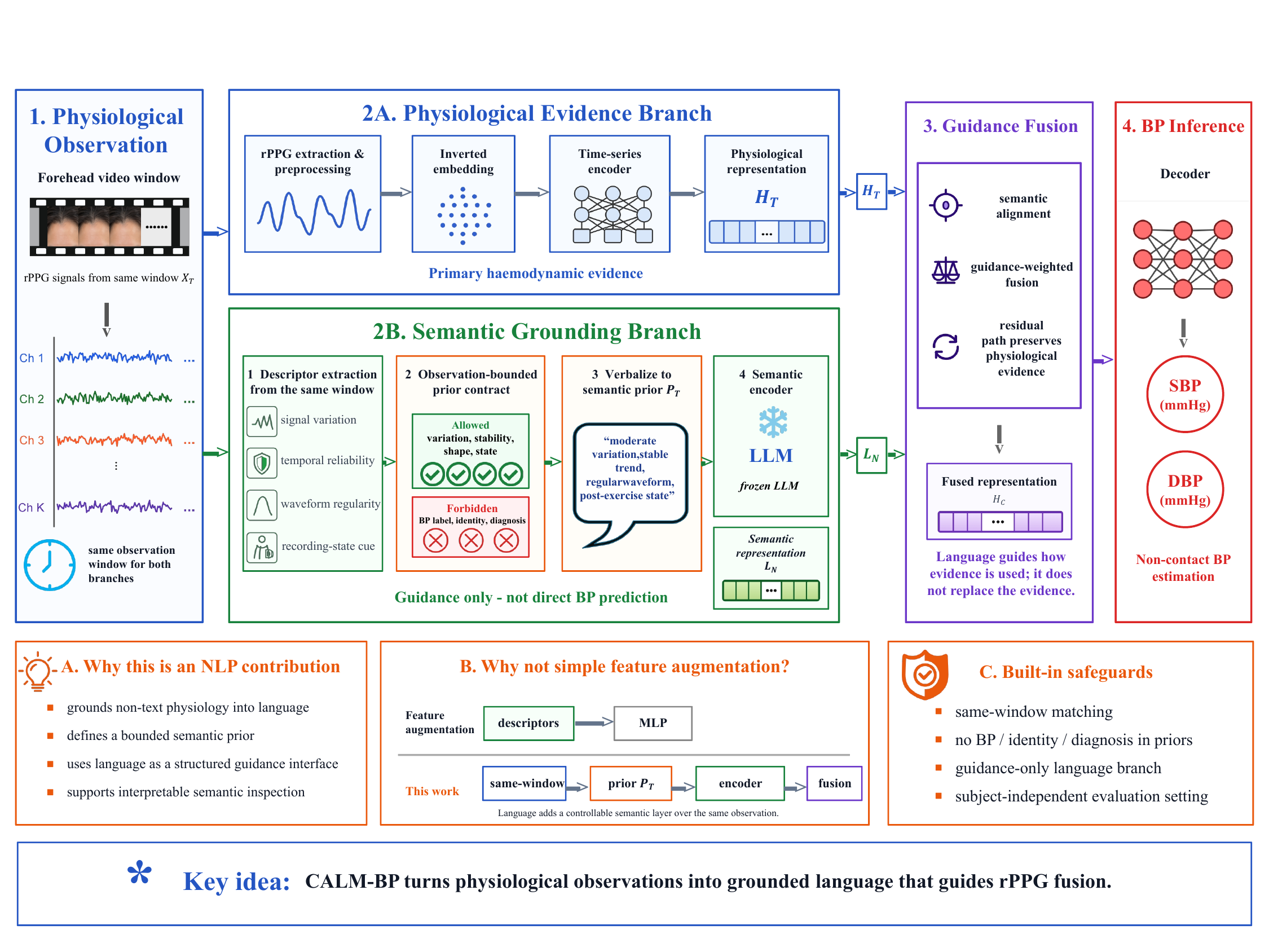}
    \caption{\textbf{Observation-matched semantic grounding in CALM-BP.} The rPPG branch provides primary physiological evidence, while a bounded semantic prior is constructed from the same observation window and encoded as guidance for cross-modal fusion. Language guides how measured evidence is used; it does not directly predict BP.}
    \label{fig:framework}
\end{figure*}

Figure~\ref{fig:framework} summarizes the evaluated CALM-BP pathway: the physiological branch encodes measured rPPG evidence, the semantic branch verbalizes observation-bounded descriptors, and the fusion module aligns the two before SBP/DBP inference.

Our contributions are as follows:
\begin{itemize}
    \item \textbf{We formulate observation-matched physiological semantic grounding} for non-contact BP estimation, where language priors organize descriptors from the same rPPG window without adding BP-label or identity information.
    \item \textbf{We propose CALM-BP,} a dual-path model that keeps numerical rPPG as primary evidence and aligns it with bounded semantic priors under an explicit prior contract.
    \item \textbf{We provide FlowBP-Set and grounding controls} for studying paired forehead observations, BP targets, promptized physiological semantics, direct cross-dataset evaluation, mismatched priors, and a leakage-audited split protocol.
\end{itemize}

\section{Related Work}

\subsection{rPPG and BP Estimation}

rPPG enables physiological information to be derived from facial video without a contact sensor \cite{Verkruysse2008rPPG,Poh2010RemotePulse}. Prior BP estimation work has modeled facial video, PPG/rPPG signals, synthetic data, or transfer learning for the prediction task \cite{Zhou2019FaceBP,Schrumpf2021PPGrPPG,Wu2022rPPGBP,Chen2023RemoteBP,Leitner2022PPGTL}. These studies motivate our signal setting. Our emphasis is different: CALM-BP keeps rPPG as the main evidence source while asking how observation-matched semantic priors should be constructed, bounded, and tested.

\subsection{Semantic Guidance for Signals}

Foundation-model and LLM-based time-series research has explored how sequence observations can be paired with broader representation spaces \cite{Liang2024FM4TS,Jiang2024LLMSurvey}. TimeCMA aligns multivariate temporal inputs with language-model features, and related studies examine LLM representations for sequence tasks \cite{Liu2024TimeCMA,Chen2025LLMFew}. SensorLLM aligns motion-sensor signals with trend descriptions, while ProMedTS uses prompt-guided learning to integrate medical text with structured time series \cite{Li2025SensorLLM,Niu2025ProMedTS}. These studies motivate semantic guidance for non-text observations. CALM-BP focuses on a physiological case where the guidance comes from promptized rPPG cues and must remain subordinate to measured haemodynamic evidence.

\subsection{LLMs and MLLMs as Interfaces}

LLMs can turn instructions and intermediate descriptions into structured semantic representations and guidance signals \cite{Brown2020LanguageModels,Wei2022ChainOfThought}. MLLMs further extend language-centered interfaces to visual inputs. For grounding, the relevant question is not whether language can directly answer from every input, but whether it can provide a bounded interface between a measured observation and a downstream task. In physiological inference, that distinction is crucial: a semantic pathway can summarize signal quality, reliability, and state-aware cues without claiming that language itself measures the body or supplies additional physiological evidence. CALM-BP adopts this view by using language-derived priors to guide fusion with a physiological branch.

\section{Task Formulation and FlowBP-Set}

We study observation-matched physiological semantic guidance for non-contact BP estimation. Each instance begins with a short forehead-video window centered on a synchronized BP timestamp. From that window we derive a measured rPPG observation \(X_T \in \mathbb{R}^{T\times K}\), where \(T\) is the number of frames and \(K=3\) denotes the RGB-derived channels. We also construct a prompt \(P_T\) that expresses physiological semantic priors for the same window. The model estimates the corresponding systolic and diastolic BP target \(Y=[\mathrm{SBP},\mathrm{DBP}]\). The two inputs have different roles: \(X_T\) carries physiological evidence, while \(P_T\) provides bounded guidance about variation, stability, waveform shape, and state-aware cues for fusion.

\subsection{FlowBP-Set}

FlowBP-Set supports this formulation with approximately 145,800 synchronized SBP/DBP observation windows and about 1.9 million retained frame-level forehead images from 81 participants. Each instance links a non-contact forehead observation, synchronized SBP and DBP targets, and a structured prompt derived from the corresponding rPPG window. The dataset covers resting, deep-breathing, and post-exercise states, allowing semantic priors to describe physiological observations under different recording states rather than a single static condition.

\subsection{Experimental Setup and Observation Processing}

FlowBP-Set was collected with a high-definition forehead camera and a CNAP reference-BP system driven by a shared clock. The camera stream was sampled at 30 FPS, and the CNAP signal was downsampled from 1000 Hz to the video frame rate. A total of 81 volunteers participated voluntarily under resting, deep-breathing, and post-exercise conditions. This setup provides paired non-contact video observations and synchronized reference SBP/DBP values for each retained physiological window.

The video stream is first converted into frame-level forehead observations. For each valid BP timestamp, we localize the forehead region and define a centered 2-second observation interval. From this interval, 21 frames are uniformly sampled from the original 30-FPS video stream, corresponding to an effective sampling rate of approximately 10 Hz. Average RGB intensities are then computed within the forehead ROI to construct the local rPPG observation. The same 2-second observation interval is used for both the numerical branch and semantic-prior construction, ensuring that the visual frames, rPPG trace, and prompt are temporally aligned with the same BP timestamp. The resulting observation is intended to characterize short-term signal variation, temporal consistency, and local waveform regularity rather than fine-grained beat morphology or long-range cardiovascular dynamics. The resulting raw chromatic trace is:
\begin{equation}
s_t = [R_t,\; G_t,\; B_t]^\top.
\end{equation}
Stacking the RGB vectors in temporal order yields a three-channel temporal sequence \(x_i(t)\), where subscript \(i\) denotes the image segment and \(t\) the frame index. Detrending, normalization, motion-artifact suppression, VMD-based component selection, and FIR denoising produce the final multivariate observation \cite{Poh2010RemotePulse,Dragomiretskiy2014VMD,deHaan2013Chrom,Wang2017AlgRemotePPG}:
\begin{equation}
X_T = [x_i(1),\, x_i(2),\, \ldots,\, x_i(T)] \in \mathbb{R}^{T \times 3},
\end{equation}
where \(T=21\) and \(K=3\). The raw trace is processed with detrending, normalization, and temporal denoising to reduce illumination drift, motion-induced fluctuation, and high-frequency noise before it enters the model. The cleaned waveform is then used both as the numerical input and as the source for semantic descriptors. Amplitude variation is computed from channel-level changes within the window, local trend from short-term temporal slope, stability from within-window fluctuation, and waveform regularity from the consistency of the cleaned rPPG shape.

We screen windows before model training or evaluation. A window is retained only when the forehead ROI is visible, the ROI position is stable, the video and CNAP streams remain synchronized, and a continuous reference BP value is available. Segments with severe motion artefacts, illumination instability, ROI drift, missing CNAP references, or boundary effects are excluded; the first and last 5 seconds of each segment are also removed to reduce resampling edge effects. 

\subsection{Semantic Prior Construction}

The semantic side of the task describes the physiological observation instead of posing a medical dialogue or generating a direct BP answer. For each rPPG window, we compute descriptors tied to observable signal behavior: channel-wise amplitude changes, local trends, stability indicators, basic waveform-shape descriptions, and recording-state cues when available. We verbalize those descriptors into a concise prompt that acts as a structured interface for the same window used by the physiological branch, not as an independent source of BP evidence.

The prompt is intentionally bounded. It can surface signal-quality, reliability, temporal, or state-aware cues, but it describes the rPPG window rather than a participant identity, a BP target, a diagnostic conclusion, or a natural-language answer. For example, a prompt may state: \emph{the RGB traces show moderate channel variation, a stable local trend, and a regular waveform shape over the current observation window}. This text is not meant to replace the rPPG sequence. It guides cross-modal inference with interpretable physiological semantics while the numerical branch retains the primary haemodynamic evidence.

The construction pipeline follows three steps. First, observation descriptors are computed from the retained physiological window and dataset context. Second, each descriptor is mapped to a bounded semantic field, such as a variation level, stability cue, waveform-shape phrase, or recording-state cue. Degree words such as low, moderate, high, stable, drifting, regular, and irregular are produced by fixed quantization rules estimated from training-participant windows only and then applied unchanged to held-out and external windows. Third, the fields are verbalized into a fixed prior template for the guidance encoder. Appendix~\ref{app:prompt_rules} gives the prompt-generation rules.

\begin{table*}[!t]
\small
\centering
\caption{\textbf{Observation-bounded semantic prior contract.} The semantic prior is restricted to descriptors derived from the current rPPG observation window and excludes target labels, identity information, and dataset-specific shortcuts.}
\label{tab:prior_contract}

\begin{tabular}{lccc}
\toprule
\textbf{Semantic field} &
\textbf{Observation basis} &
\textbf{Verbalized prior content} &
\textbf{Grounding role} \\
\midrule

Signal variation &
\shortstack[c]{RGB/rPPG amplitude\\fluctuation} &
\shortstack[c]{low, moderate, or high\\variation} &
\shortstack[c]{variation-aware\\guidance} \\

Temporal reliability &
\shortstack[c]{local trend\\consistency} &
\shortstack[c]{stable or drifting\\trend} &
\shortstack[c]{reliability-aware\\fusion} \\

Waveform regularity &
\shortstack[c]{shape and\\morphology cues} &
\shortstack[c]{regular or irregular\\waveform} &
\shortstack[c]{waveform-level\\guidance} \\

Recording-state cue &
\shortstack[c]{same-window\\protocol state} &
\shortstack[c]{resting, breathing,\\or post-exercise} &
\shortstack[c]{bounded recording\\context} \\

\midrule

Forbidden information &
\shortstack[c]{not permitted\\in the prior} &
\shortstack[c]{BP, identity, diagnosis,\\demographics, split} &
\shortstack[c]{prevents leakage\\and shortcuts} \\

\bottomrule
\end{tabular}

\end{table*}

\input{t1}

Table~\ref{tab:prior_contract} specifies the semantic prior contract used throughout the paper. Each prior is derived only from the current rPPG observation window. The contract separates what the semantic pathway may summarize from what must remain in the physiological inference path, ensuring that language acts as semantic grounding guidance rather than additional physiological evidence or a hidden label channel. This distinction is central to the formulation: semantic priors can guide reliability-aware fusion, but they cannot introduce BP targets, describe participant identity, or convert appearance into a direct BP answer.

\begin{figure*}[!t]
    \centering
    \includegraphics[width=\textwidth]{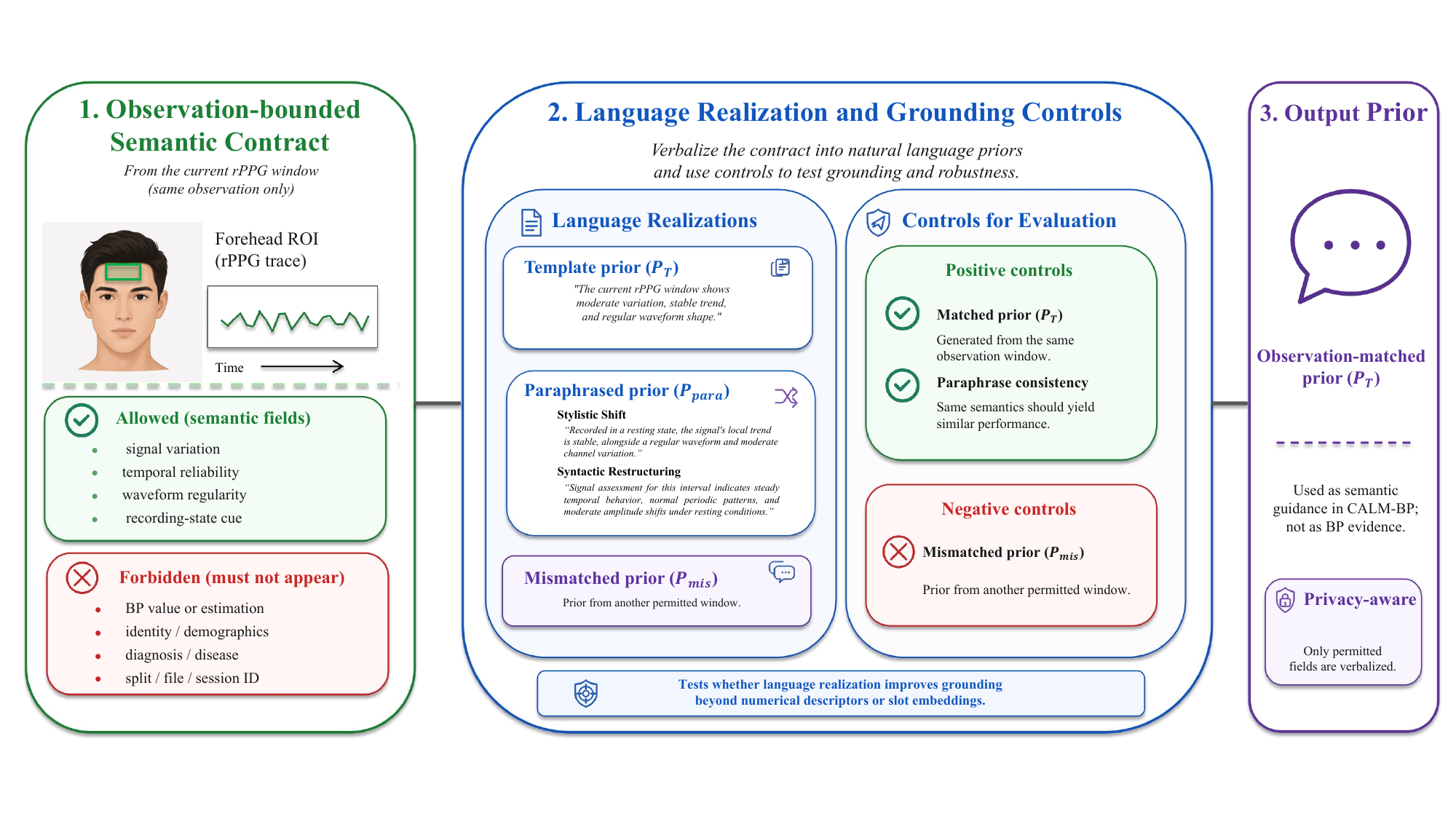}
    \caption{\textbf{Observation-to-language grounding prior construction.}
    CALM-BP converts same-window rPPG descriptors into a bounded semantic prior. The permitted fields are verbalized as the matched prior $P_T$, paraphrased prior $P_{\mathrm{para}}$, or mismatched prior $P_{\mathrm{mis}}$ for grounding-control evaluation. The prior guides fusion with rPPG evidence and is not used as independent BP evidence.}
    \label{fig:prior_trace}
\end{figure*}

\section{CALM-BP}

\subsection{Overview}

CALM-BP instantiates the pathway summarized in Figure~\ref{fig:framework}. A physiological signal branch encodes rPPG dynamics from the forehead observation. A semantic guidance layer encodes promptized priors about the same window. Cross-modal fusion then lets the physiological representation consult those priors before BP inference. We use \(K=3\) for the RGB-derived rPPG channels, \(H_T\) for the physiological representation, \(L_N\) for the language-derived semantic representation, and \(H_C\) for the fused representation. The separation is deliberate: semantic priors guide the prediction pathway, but they do not replace measured physiological evidence.

\subsection{Semantic Instruction Construction}

In CALM-BP, semantic guidance is defined by an intermediate guidance state rather than by direct text generation of BP. Figure~\ref{fig:prior_trace} illustrates this trace from same-window rPPG descriptors to bounded semantic fields and then to the language prior. For an observation window, descriptor extraction yields a bounded semantic record \(D_T\) containing variation, temporal, shape, and state cues. A verbalization operator turns that record into the physiological prior \(P_T\):
\begin{equation}
    P_T = \operatorname{Verbalize}(D_T).
\end{equation}
The semantic guidance layer encodes \(P_T\) into \(L_N\), which is available to the physiological branch during fusion. The pathway separates observation summarization, prior construction, and downstream inference, making it a bounded language interface over non-text physiological observations. Its novelty is the observation-matched contract: the prior must be derived from the current physiological window and must not predict BP from text or appearance alone.

\subsection{Physiological Branch}

The physiological signal branch is the main evidence pathway. For the three-channel rPPG sequence \(X_T\), we first apply an inverted embedding strategy that treats the multivariate sequence as a tokenized temporal observation. A linear projection produces the initial embedding:
\begin{equation}
H_T = W_e X_T + b_e,
\end{equation}
where \(W_e\) and \(b_e\) are learnable parameters. The embedded sequence is then passed into a lightweight Pre-LN Transformer. This branch preserves channel interaction and local temporal variation that semantic guidance cannot recover after the fact.

\subsection{Language Prior Encoder}

The semantic guidance layer converts structured physiological priors into language-model features. Physiological prompts are tokenized and fed into a frozen Qwen3-0.6B \cite{QwenTeam2025Qwen3} model. We extract its final-token representation and pass it through a prompt encoder that mirrors the physiological encoder. The result \(L_N\) is not a free-form BP inference. It is the encoded guidance representation of \(P_T\), designed to meet the physiological embedding \(H_T\) during fusion.

\subsection{Fusion and BP Inference}

Semantic-guided inference requires the priors to interact with the physiological evidence instead of being appended as an undifferentiated auxiliary vector. We project \(H_T\) and \(L_N\) into a comparable representation space and compute a channel-wise similarity matrix. The matrix lets the physiological branch retrieve observation-matched guidance:
\begin{equation}
M_T = \sigma \left( \psi_q(H_T) \otimes \psi_k(L_N) \right),
\end{equation}
where \(\sigma(\cdot)\) denotes softmax normalization and \(\psi_q(\cdot)\) and \(\psi_k(\cdot)\) are learnable linear mappings. We aggregate the textual representations according to the similarity weights and fuse them with the numerical embedding through a residual connection:
\begin{equation}
H_C = \omega_c \left( \psi_v(L_N) \otimes M_T \right) + H_T,
\end{equation}
where \(\psi_v(\cdot)\) is another linear projection and \(\omega_c(\cdot)\) denotes the fusion layer. The residual form keeps the physiological path available after fusion, so \(H_C\) incorporates semantic priors while remaining anchored to rPPG structure.

The fused representation \(H_C\) is fed into a multivariate Transformer decoder that further models dependencies among the three RGB channels. The decoder output \(\hat{H}_C\) is projected into the BP space:
\begin{equation}
    \hat{Y} = W_p \hat{H}_C + b_p,
\end{equation}
where \(W_p\) and \(b_p\) are learnable projection parameters, and \(\hat{Y}\) contains the predicted SBP and DBP values for the window.

\subsection{Loss Function}

The model minimizes a mean squared error objective with L2 regularization:
\begin{equation}
\mathcal{L} = \| \hat{Y} - Y \|_2^2 + \lambda \| W \|_2^2,
\end{equation}
where \(W\) denotes learnable parameters and \(\lambda\) is the regularization coefficient. We set \(\lambda=10^{-4}\) in all experiments. After inverted embedding projection, the numerical and text branches produce latent representations of dimensions \(C=128\) and \(E=128\).

\section{Experiments}

We evaluate whether language-derived physiological priors help only when paired with the current rPPG observation. Section~5.1 gives the setup, Section~5.2 reports FlowBP-Set results, Section~5.3 tests direct cross-dataset transfer, and Section~5.4 studies language realization and observation-matching controls.

\input{t3}

\subsection{Experimental Setup}

\subsubsection{Datasets}
FlowBP-Set contains 81 participants, approximately 145,800 synchronized SBP/DBP windows, and about 1.9 million retained forehead images. We use a 65/16 participant-disjoint train/test split created before window sampling; normalization uses training participants only, and prompts exclude SBP, DBP, identity, diagnosis, and demographic identifiers.

\subsubsection{Metrics}
For BP estimation, we report mean absolute error (MAE), root mean squared error (RMSE), and standard deviation of error (SD) for SBP and DBP.

\subsubsection{Implementation Details}
We train in PyTorch with batch size 4 and learning rate \(1\times10^{-6}\) on one NVIDIA RTX A100 GPU. The Qwen3-0.6B backbone is frozen; the rPPG encoder, prompt encoder, alignment block, Transformer decoder, and regression head are learned. All normalization statistics are estimated from training participants only.

\subsection{Main Results}

Table~\ref{tab1} compares FlowBP-Set results across resting, deep-breathing, and post-exercise states. This section reports only the in-domain subject-disjoint setting on our own dataset. CALM-BP improves over rPPG and BP baselines while retaining rPPG as the primary physiological evidence source.

\subsection{Cross-Dataset Transfer}

To test direct generalization, we train CALM-BP on FlowBP-Set and evaluate the fixed model on the BESTLab BP benchmark associated with BPNet \cite{Manullang2025TransformerBP} and Vital Videos (VV) \cite{Toye2023VitalVideos}. External rPPG observations and semantic priors are reconstructed with the same preprocessing and descriptor-to-prompt pipeline; no external BP labels are used to train or fine-tune CALM-BP.

Tables~\ref{tab2} and~\ref{tab3} evaluate the FlowBP-Set-trained model directly on the BESTLab BP benchmark and VV. These results test whether the learned physiological and semantic interface transfers beyond the collected FlowBP-Set distribution without external BP-label adaptation.

\subsection{Ablation Studies}

Tables~\ref{tab4}, \ref{tab:semantic_content_controls}, and \ref{tab:prompt_surface_robustness} report metrics on the resting-state subset, while Tables~\ref{tab:language_realization_ablation} and \ref{tab:observation_matching_controls} report window-count-weighted overall metrics across the resting, deep-breathing, and post-exercise states.

\input{t2}

Table~\ref{tab4} isolates the contribution of each pathway component under the same Qwen3-0.6B backbone. Removing the time-series encoder, the tokenizer, or the prompt encoder degrades performance progressively, while the full configuration with the frozen LLM achieves the best result. This confirms that the gain stems from the combination of a dedicated physiological encoder and a language-derived prior encoder rather than from any single component alone.

\input{t11}

Table~\ref{tab:language_realization_ablation} compares direct numerical descriptors, discrete slot embeddings, and verbalized language priors under the same rPPG branch. The comparison tests whether language realization adds value beyond feature augmentation.

\input{t7}

Table~\ref{tab:semantic_content_controls} verifies that performance depends on observation-matched physiological content rather than the mere presence of a language branch. Weakening the prior, using generic template text, or using random non-task text degrades results, and a contradictory prior is also worse than the full prior, indicating that the prior must describe the actual observation.

\input{t9}

Table~\ref{tab:prompt_surface_robustness} shows that CALM-BP is robust to prompt surface form. Paraphrasing, shuffling field order, replacing synonyms, and using multiple templates all preserve performance close to the original prompt, so the model does not rely on fixed prompt positions or surface templates.

\input{t4}

Table~\ref{tab:observation_matching_controls} tests surface robustness and grounding. The paraphrased matched prior keeps the same observation pairing but changes wording, while mismatch rows break prior--observation pairing at random, within-subject, or state-preserved levels. The degradations show that neither arbitrary text, subject identity, nor coarse state similarity replaces window-level semantic alignment.

\subsection{Result Discussion}

Overall, CALM-BP improves on the subject-disjoint FlowBP-Set split, transfers directly to external targets, and benefits most when the semantic prior remains matched to the current observation.

\section{Discussion}

CALM-BP investigates whether observation-derived semantic guidance can improve non-contact BP estimation while preserving rPPG as the primary physiological evidence source. Rather than introducing additional physiological measurements, the semantic pathway summarizes selected properties of the same rPPG observation, including signal variation, temporal reliability, local waveform regularity, and recording state. These descriptors are verbalized under an explicit prior contract and incorporated through cross-modal fusion, providing a structured and auditable mechanism for conditioning BP estimation on the quality and characteristics of the current physiological observation.

The experimental results suggest that the benefit of the semantic pathway is not explained solely by adding auxiliary descriptors. Under matched information content, verbalized semantic priors encoded by the frozen language model outperform the evaluated numerical-descriptor and categorical-embedding alternatives. Importantly, this result should be interpreted within the evaluated representation settings rather than as evidence that natural language is universally superior to numerical features. The observation-mismatch experiments provide complementary evidence: performance degrades when the semantic prior is detached from its corresponding rPPG window, including under within-subject and state-preserved mismatches. This indicates that the improvement depends on observation-level correspondence between physiological evidence and semantic guidance rather than on arbitrary textual input, participant-specific similarity, or coarse recording-state information alone.

The subject-disjoint evaluation and direct cross-dataset experiments further examine whether this formulation generalizes beyond the training observations. Although CALM-BP shows improved performance under the evaluated settings, its effectiveness remains dependent on the reliability of the underlying rPPG signal and the descriptors derived from it. The current semantic prior represents a predefined subset of short-term physiological characteristics and therefore cannot recover information that is absent or severely corrupted in the measured signal. These findings position CALM-BP as a semantic-guided physiological signal estimation framework rather than an independent language-based BP predictor, and motivate future work on more robust rPPG extraction, broader population and acquisition conditions, longer temporal modeling, and alternative semantic encoders.

\section{Conclusion}

We present FlowBP-Set and CALM-BP for observation-matched physiological semantic grounding in non-contact BP estimation. CALM-BP keeps rPPG as the primary evidence source and uses same-window semantic priors to guide fusion; results on FlowBP-Set, direct external transfer, and grounding controls show that the benefit comes from matched semantic guidance rather than arbitrary text or label leakage. Future work should broaden population coverage, semantic perturbation tests, and downstream health-language validation.

\section*{Limitations}

CALM-BP studies observation-matched semantic guidance for one physiological target, BP estimation from rPPG windows and promptized semantic priors. The current work focuses on a bounded grounding interface between measured physiological evidence and language-derived priors; it does not instantiate a complete health-language system with planning, memory update, or tool orchestration. The prompt semantics depend on a predefined descriptor construction process and may omit physiological cues not captured by the selected descriptors. The main-paper controls test observation matching under random, within-subject, and state-preserved mismatches, but they do not exhaustively perturb every individual semantic field or prompt surface form. CALM-BP predicts SBP and DBP with interpretable semantic cues rather than a clinically validated continuous waveform, and it should not be read as a clinical decision system or deployment-ready monitor without appropriate validation, calibration, and human oversight.

\section*{Ethical Considerations}

FlowBP-Set is based on physiological recordings from human participants. The study protocol was reviewed and participation was voluntary. Because facial video, physiological traces, and BP labels can expose sensitive health information, any release should document consent, privacy handling, access scope, and intended use. Promptized descriptors and semantic guidance summaries should be handled with the same care when they remain linked to physiological windows. The paper positions CALM-BP as a semantic-guidance estimation method rather than a diagnostic system, and downstream health uses should account for measurement error, population coverage, and the risks of over-interpreting non-contact estimates.

\section*{CRediT authorship contribution statement}
\textbf{Sun Haiyang:} Conceptualization, Methodology, Software, Writing -- original draft. \textbf{Gu Boyuan:} Data curation, Validation, Writing -- review \& editing. \textbf{Liu Yongjie:} Supervision, Funding acquisition.

\section*{Declaration of Competing Interest}
The authors declare that they have no known competing financial interests or personal relationships that could have appeared to influence the work reported in this paper.

\section*{Acknowledgements}
This work was supported by [Funding Agency, Grant No. XXX]. The authors thank the participants of FlowBP-Set for their voluntary participation.

\appendix

\section{Additional FlowBP-Set Details}
\label{app:flowbp}

This appendix provides additional details on FlowBP-Set, including cohort composition, acquisition synchronization, window construction, quality screening, and privacy-aware dataset reporting.

FlowBP-Set contains 81 participants, approximately 145,800 synchronized SBP/DBP windows, and about 1.9 million retained forehead images. The main split uses 65 training participants and 16 held-out participants. All windows from one participant remain in a single split, and splits are created before window sampling. Table~\ref{tab:protocol_summary} summarizes how FlowBP-Set and the two external target datasets are used in evaluation.

\input{t5}

\subsection{Acquisition and Synchronization}
\label{app:flowbp_acquisition}

FlowBP-Set was collected using a high-definition forehead-facing camera and a CNAP reference-BP system synchronized by a shared acquisition clock. The camera stream was sampled at 30 FPS, and the CNAP reference signal was downsampled from 1000 Hz to the video frame rate. This produces a synchronized timestamp sequence in which each retained BP timestamp can be paired with a temporally aligned forehead observation window.

The recording protocol includes resting, deep-breathing, and post-exercise states. These states are used only as bounded recording-context cues when they are synchronized with the same observation window. They are not converted into BP categories and are not used to infer participant identity or diagnostic status.

\subsection{Window Quality Screening}
\label{app:window_quality}

The retained windows follow the same screening rules used in the main experiments. We require a visible and stable forehead ROI, synchronized video and CNAP timestamps, and a continuous reference BP value for the full observation window. Windows are excluded when severe motion, illumination instability, ROI drift, missing reference values, or segment-boundary effects would make the observation unreliable. Normalization statistics are estimated only from training participants and then fixed for held-out participants and external target datasets.

\subsection{Observation-Mismatch Control Construction}
\label{app:mismatch_controls}

The main grounding question is whether the semantic prior must match the current physiological observation. We therefore construct mismatch controls by changing the pairing between $X_T$ and $P_T$ while keeping the target BP label attached to the original physiological observation. In other words, the model receives the original rPPG window and predicts the original SBP/DBP target, but the semantic prior is replaced by a prior generated from another permitted window.

Let $(X_i, P_i, Y_i)$ denote the original sample, where $X_i$ is the rPPG observation, $P_i$ is the semantic prior generated from $X_i$, and $Y_i$ is the synchronized BP target. The matched setting uses $(X_i, P_i, Y_i)$. A mismatched control constructs $(X_i, P_j, Y_i)$ with $j \neq i$. The BP target is never changed during mismatch construction.

We use three mismatch settings:
\begin{enumerate}
    \item \textbf{Random mismatch:} $P_j$ is sampled from another retained window without constraining participant grouping or recording state. This tests whether arbitrary auxiliary text can improve the model when the prior is detached from the current observation.

    \item \textbf{Within-subject mismatch:} $P_j$ is sampled from a different window of the same participant. This preserves subject-level sampling similarity while breaking exact window-level observation matching. No explicit identity field is exposed to the semantic prior.

    \item \textbf{State-preserved mismatch:} $P_j$ is sampled from another window with the same recording state. This preserves the coarse resting, deep-breathing, or post-exercise cue while breaking the exact observation-prior pairing.
\end{enumerate}

\subsection{Difference Between Laboratory and In-the-Wild External Evaluation}
\label{app:external_difference}

Tables~3 and~4 show different transfer behavior on the BESTLab Physio BP benchmark and VV. BESTLab Physio represents a more controlled laboratory setting, whereas VV is closer to an in-the-wild video benchmark with broader capture variability. This difference may explain why CALM-BP obtains lower errors on BESTLab Physio than on VV, especially for SBP estimation.

\paragraph{BESTLab Physio availability.}
BESTLab Physio was used from a previously obtained local copy. At the time of writing, we could not identify a stable public download link for this dataset. We therefore do not redistribute BESTLab Physio and report it only as an external evaluation target. The preprocessing, descriptor construction, and evaluation protocol used in our experiments are described to support transparency.

In the controlled laboratory setting, face position, illumination, camera distance, and participant behavior are usually more stable. These conditions make forehead ROI tracking, RGB/rPPG extraction, and semantic descriptor construction more reliable. As a result, the observation-matched prior is more likely to describe the same physiological evidence seen by the numerical branch.

By contrast, in-the-wild videos may contain larger changes in pose, lighting, motion, compression, skin-region visibility, and background conditions. These factors can reduce rPPG quality and make descriptors such as signal variation, temporal reliability, and waveform regularity less stable. In addition, recording-state cues may be unavailable or less standardized in external datasets, so the semantic prior cannot always use the same contextual information as in FlowBP-Set.

Therefore, the gap between Table~3 and Table~4 should be interpreted as evidence that cross-dataset generalization is sensitive to capture conditions and signal quality. CALM-BP transfers without target-label fine-tuning, but its performance still depends on whether the external video provides a stable physiological observation window. This suggests that improving robust ROI tracking, illumination normalization, and longer temporal modeling is important for future in-the-wild non-contact BP estimation.

\section{Leakage and Split Audit}
\label{app:leakage_audit}

Table~\ref{tab:leakage_audit} summarizes the split and prompt safeguards used by the main experiments. The evaluation design is intentionally conservative: the physiological branch remains responsible for BP inference, while the semantic branch is restricted to bounded same-window descriptors. If a semantic prior is detached from its source observation, the mismatch controls test whether the benefit persists.

\input{t8}

\section{Semantic Prior Construction and Prompt Audit}
\label{app:prompt_rules}

This appendix specifies how the semantic prior $P_T$ is constructed from the current physiological observation $X_T$. The goal is not to convert rPPG into a free-form medical interpretation, but to expose a bounded and auditable language interface over observable signal behavior. Each prior is generated deterministically from the same rPPG window used by the physiological branch, with a fixed set of allowed semantic fields and an explicit list of forbidden information.

\begin{table*}[!t]
\footnotesize
\renewcommand{\arraystretch}{0.95}
\centering
\caption{\textbf{Descriptor-to-semantic-field mapping.} The semantic prior is generated from bounded observation-level descriptors used by the semantic grounding branch. Quantization thresholds are estimated from training-participant windows only, fixed before evaluation, and applied unchanged to held-out and external windows. The mapping is deterministic and does not use BP values, participant identity, diagnosis, demographics, split membership, file paths, or session IDs.}
\label{tab:descriptor_mapping}
\resizebox{\textwidth}{!}{%
\begin{tabular}{p{0.18\textwidth} p{0.27\textwidth} p{0.31\textwidth} p{0.18\textwidth}}
\toprule
\textbf{Semantic field} & \textbf{Observation basis} & \textbf{Quantization rule} & \textbf{Verbalized label} \\
\midrule
Signal variation
& RMS fluctuation over the three preprocessed RGB/rPPG channels
& $V_{\mathrm{amp}} \le q_{33}^{V}$; $q_{33}^{V} < V_{\mathrm{amp}} \le q_{67}^{V}$; $V_{\mathrm{amp}} > q_{67}^{V}$, where $q_{33}^{V}$ and $q_{67}^{V}$ are computed from training-participant windows
& low, moderate, or high signal variation \\

Temporal reliability
& Absolute least-squares slope of the channel-mean sequence within the current window
& $|k_{\mathrm{trend}}| \le q_{75}^{K}$ or $|k_{\mathrm{trend}}| > q_{75}^{K}$, where $q_{75}^{K}$ is computed from training-participant windows
& stable or drifting local trend \\

Waveform regularity
& VMD-derived noise energy ratio between pulse-like and residual high-frequency components
& $NER \le q_{75}^{N}$ or $NER > q_{75}^{N}$, where $q_{75}^{N}$ is computed from training-participant windows
& regular or irregular waveform \\

Recording-state cue
& Synchronized protocol log for the same observation window
& Included only when available for the current window
& resting, deep-breathing, or post-exercise state \\

Forbidden information
& Not permitted in semantic-prior construction
& Never read by the prompt generator
& SBP, DBP, BP category, identity, diagnosis, demographics, split membership, file path, or session ID \\
\bottomrule
\end{tabular}
}
\end{table*}

\subsection{Overview of Prior Construction}
\label{app:prompt_overview}

For each retained observation window, semantic-prior generation contains three steps:
\begin{enumerate}
    \item Compute observation-level descriptors from the preprocessed rPPG window $X_T$.
    \item Quantize each descriptor into a bounded semantic field.
    \item Populate a fixed verbalization template using only allowed fields.
\end{enumerate}
After the prior is generated, CALM-BP encodes the resulting prompt with the frozen semantic encoder during model inference and training. We separate prior generation from model-side encoding so that the prompt-construction protocol remains auditable and independent of the downstream architecture.

The prompt generator never reads SBP, DBP, participant identity, diagnosis, demographic fields, split membership, file paths, session IDs, or target-derived BP categories. The resulting text is therefore a semantic description of the current observation window, not an independent BP measurement or a hidden label channel.

\paragraph{Rationale for deterministic priors.}
We intentionally use deterministic prior construction rather than free-form LLM generation, making the semantic pathway auditable and reproducible. The thresholds used for descriptor quantization are fixed before evaluation and are applied unchanged to held-out participants and external datasets. All normalization statistics used by the signal-processing pipeline are estimated from training participants only. Each token in the prior can be traced to a predefined observation-level descriptor or a synchronized recording-state cue. This prevents the language branch from introducing uncontrolled medical interpretations, dataset-specific hints, or latent BP-related assumptions. In CALM-BP, language does not generate new physiological evidence or freely reason about BP; it provides a controlled semantic realization of the same rPPG observation. This also makes the grounding claim testable: paraphrased matched priors should remain stable, whereas mismatched priors should degrade performance.

\subsection{Descriptor-to-Semantic-Field Mapping}
\label{app:descriptor_mapping}

Table~\ref{tab:descriptor_mapping} summarizes the descriptor-to-field mapping used for semantic-prior construction. All fields are derived from the current rPPG window, except the recording-state cue, which is included only when it is available from the synchronized protocol log for the same observation window.

\subsection{Field Definitions}
\label{app:field_definitions}

\paragraph{Signal variation.}
For the preprocessed three-channel signal $X_T \in \mathbb{R}^{21 \times 3}$, we calculate the root mean square fluctuation across the channels:
\begin{equation}
V_{\mathrm{amp}} = \sqrt{\frac{1}{3} \sum_{c \in \{R,G,B\}} \sigma_c^2}.
\end{equation}
Let $q_{33}^{V}$ and $q_{67}^{V}$ denote the 33rd and 67th percentiles of $V_{\mathrm{amp}}$ computed from training-participant windows only. The descriptor is verbalized as ``low signal variation'' when $V_{\mathrm{amp}} \le q_{33}^{V}$, ``moderate signal variation'' when $q_{33}^{V} < V_{\mathrm{amp}} \le q_{67}^{V}$, and ``high signal variation'' when $V_{\mathrm{amp}} > q_{67}^{V}$. These thresholds are fixed before evaluation and applied unchanged to held-out and external windows.

\paragraph{Temporal reliability.}
To characterize short-window temporal reliability, we apply a least-squares linear fit to the channel-mean sequence. The resulting slope $k_{\mathrm{trend}}$ is mapped to ``stable trend'' when $|k_{\mathrm{trend}}| \le 0.02$ and ``drifting trend'' when $|k_{\mathrm{trend}}| > 0.02$. This field describes local signal behavior rather than BP direction or BP magnitude.

\paragraph{Waveform regularity.}
Waveform morphology is quantized using a high-frequency Noise Energy Ratio derived from Variational Mode Decomposition. Let $E_{\mathrm{pulse}}$ be the energy of the core pulse-like mode and $E_{\mathrm{noise}}$ be the residual high-frequency energy. We define
\begin{equation}
NER = \frac{E_{\mathrm{noise}}}{E_{\mathrm{pulse}} + E_{\mathrm{noise}}}.
\end{equation}
Windows are labeled ``regular waveform'' if $NER < 0.2$ and ``irregular waveform'' otherwise. This field provides a reliability-oriented waveform description and is not used to infer a BP label directly.

\paragraph{Recording-state cue.}
The state field is drawn only from synchronized protocol logs for the same observation window. It may take values such as resting, deep-breathing, or post-exercise. When the state field is unavailable, as may occur in external datasets, it is omitted rather than imputed. The state cue is included as bounded recording context and is separately audited through the state-preserved mismatch control described in Appendix~\ref{app:mismatch_controls}.

\subsection{Mismatch Prior Construction}
\label{app:prompt_mismatch_rules}

Mismatched priors are constructed by replacing the semantic prior $P_i$ of the current observation $X_i$ with a prior $P_j$ generated from another retained window $X_j$, while keeping the physiological observation $X_i$ and BP target $Y_i$ unchanged. This construction ensures that performance changes reflect broken observation-prior matching rather than altered BP labels.

For within-subject mismatch, $X_j$ is sampled from the same participant but a different window, preserving subject-level sampling similarity without exposing identity information to the prompt. For state-preserved mismatch, $X_j$ is sampled from another window with the same recording state. For random mismatch, $X_j$ is sampled without constraining participant grouping or recording state. These settings are defined in Appendix~\ref{app:mismatch_controls} and reported in the main observation-prior grounding controls.

\bibliographystyle{unsrt}
\bibliography{refs}

\end{document}

%% file: t1.tex
\begin{table*}[htbp]
\caption{FlowBP-Set comparison for observation-matched physiological semantic grounding. CALM-BP fuses rPPG evidence with semantic priors derived from the same observation window.}
\centering
\resizebox{\textwidth}{!}{%
\begin{tabular}{clccc cccccc cccccc ccc}
\toprule
\multicolumn{2}{c}{\multirow{3}{*}{\centering\textbf{\raisebox{-3.0ex}{Method}}}} &
\multicolumn{6}{c}{\textbf{Resting State}} &
\multicolumn{6}{c}{\textbf{Deep Breath}} &
\multicolumn{6}{c}{\textbf{Exercise}} \\
\cmidrule(lr){3-20}
\multicolumn{2}{c}{} &
\multicolumn{3}{c}{\textbf{SBP}} &
\multicolumn{3}{c}{\textbf{DBP}} &
\multicolumn{3}{c}{\textbf{SBP}} &
\multicolumn{3}{c}{\textbf{DBP}} &
\multicolumn{3}{c}{\textbf{SBP}} &
\multicolumn{3}{c}{\textbf{DBP}} \\
\cmidrule(lr){3-5} \cmidrule(lr){6-8} \cmidrule(lr){9-11} \cmidrule(lr){12-14} \cmidrule(lr){15-17} \cmidrule(lr){18-20}
\multicolumn{2}{c}{} &
MAE$\downarrow$ & RMSE$\downarrow$ & SD$\downarrow$ &
MAE$\downarrow$ & RMSE$\downarrow$ & SD$\downarrow$ &
MAE$\downarrow$ & RMSE$\downarrow$ & SD$\downarrow$ &
MAE$\downarrow$ & RMSE$\downarrow$ & SD$\downarrow$ &
MAE$\downarrow$ & RMSE$\downarrow$ & SD$\downarrow$ &
MAE$\downarrow$ & RMSE$\downarrow$ & SD$\downarrow$ \\
\midrule
\multirow{6}{*}{\centering\textbf{\raisebox{4.0ex}{LLM}}}
& InternVL3.5-2B\cite{Wang2025InternVL35} & 14.96 & 21.05 & 13.80 & 10.78 & 19.92 & 10.82 & 17.79 & 21.57 & 17.05 & 16.09 & 19.87 & 15.56 & 12.80 & 19.44 & 15.73 & 16.80 & 17.90 & 13.32 \\
& mplug-owl-llama-7B\cite{Ye2023mPLUGOwl} & 13.43 & 21.22 & 10.48 & 14.69 & 17.86 & 15.16 & 19.83 & 21.06 & 14.43 & 16.52 & 19.86 & 18.88 & 16.13 & 18.75 & 17.18 & 15.96 & 20.77 & 15.12 \\
& InternVL2-8B\cite{Chen2024InternVL2} & 16.29 & 24.03 & 15.72 & 13.78 & 18.73 & 14.42 & 19.37 & 22.18 & 18.38 & 17.53 & 18.27 & 12.88 & 12.07 & 17.09 & 19.24 & 16.59 & 18.65 & 16.93 \\
& Internlm-Xcomposer2-7B-chat\cite{Dong2024XComposer2} & 14.01 & 18.21 & 12.83 & 15.07 & 21.03 & 13.49 & 14.93 & 18.97 & 16.74 & 13.26 & 18.94 & 12.74 & 14.92 & 18.53 & 14.52 & 14.18 & 17.82 & 21.98 \\
& LLaVA-v1.5-13B\cite{Liu2024LLaVA15} & 17.44 & 19.87 & 19.23 & 10.75 & 24.58 & 12.28 & 12.25 & 18.13 & 12.38 & 19.85 & 22.72 & 14.30 & 12.21 & 17.74 & 11.55 & 13.88 & 19.33 & 13.25 \\
& Qwen2.5-3B\cite{Qwen2025Qwen25} & 14.37 & 19.76 & 16.08 & 11.94 & 21.47 & 14.60 & 18.29 & 20.52 & 16.55 & 12.44 & 21.79 & 17.26 & 13.51 & 18.91 & 14.29 & 15.57 & 18.52 & 11.03 \\
\midrule
\multirow{8}{*}{\centering\textbf{\raisebox{2.5ex}{Baseline}}}
& MeanRegressor\cite{Schrumpf2021PPGrPPG} & 11.42 & 12.38 & 9.45 & 10.91 & 11.53 & 10.88 & 12.64 & 13.88 & 11.52 & 11.57 & 12.17 & 9.92 & 10.16 & 12.33 & 10.49 & 12.84 & 13.58 & 12.58 \\
& Zhou et al.\cite{Zhou2019FaceBP} & 11.23 & 12.64 & 10.66 & 9.30 & 12.67 & 9.85 & 11.70 & 12.76 & 12.54 & 12.01 & 13.14 & 10.66 & 11.88 & 12.73 & 10.15 & 9.87 & 10.30 & 9.20 \\
& Schrumpf et al. (A)\cite{Schrumpf2021PPGrPPG} & 12.87 & 13.79 & 11.21 & 11.78 & 11.79 & 10.12 & 10.85 & 12.51 & 11.36 & 11.76 & 12.79 & 12.64 & 11.54 & 12.04 & 9.54 & 12.82 & 13.57 & 10.13 \\
& Schrumpf et al. (R)\cite{Schrumpf2021PPGrPPG} & 11.59 & 12.23 & 9.07 & 10.33 & 12.02 & 11.08 & 11.17 & 12.90 & 12.12 & 13.47 & 14.31 & 12.41 & 11.41 & 12.18 & 9.41 & 9.94 & 10.64 & 11.94 \\
& BPNet-base\cite{Manullang2025TransformerBP} & 13.16 & 14.06 & 10.18 & 9.59 & 12.38 & 9.95 & 11.13 & 12.37 & 9.86 & 12.57 & 13.16 & 12.10 & 9.95 & 11.06 & 11.23 & 9.83 & 11.73 & 9.21 \\
& BPNet-medium\cite{Manullang2025TransformerBP} & 10.53 & 12.91 & 11.39 & 11.06 & 12.04 & 10.63 & 11.37 & 11.44 & 10.16 & 12.55 & 13.29 & 11.76 & 11.02 & 12.57 & 9.57 & 10.56 & 12.78 & 9.11 \\
& BPNet-large\cite{Manullang2025TransformerBP} & 9.44 & 11.92 & 9.62 & 10.93 & 11.51 & 9.40 & 10.33 & 11.63 & 9.74 & 11.12 & 12.15 & 11.08 & 10.35 & 10.73 & 9.32 & 10.77 & 12.43 & 9.31 \\
\midrule
\rowcolor{gray!25}
& \textbf{CALM-BP (ours)} & \textbf{7.94} & \textbf{9.19} & \textbf{8.37} & \textbf{8.52} & \textbf{8.91} & \textbf{7.98} & \textbf{9.83} & \textbf{10.41} & \textbf{9.18} & \textbf{10.26} & \textbf{10.92} & \textbf{9.17} & \textbf{8.73} & \textbf{8.92} & \textbf{8.91} & \textbf{9.62} & \textbf{10.18} & \textbf{8.92} \\
\bottomrule
\end{tabular}%
}
\label{tab1}
\end{table*}

%% file: t3.tex
\begin{table}[htbp]
\caption{Cross-dataset task comparison for observation-matched CALM-BP on the BESTLab dataset.}
\label{tab2}
\resizebox{\columnwidth}{!}{%
\begin{tabular}{lcccccc}
\toprule
\multicolumn{1}{c}{\multirow{2}{*}[-0.5ex]{Methods}} & \multicolumn{3}{c}{SBP} & \multicolumn{3}{c}{DBP} \\
\cmidrule(lr){2-4} \cmidrule(lr){5-7}
 & MAE$\downarrow$ & RMSE$\downarrow$ & SD$\downarrow$ & MAE$\downarrow$ & RMSE$\downarrow$ & SD$\downarrow$ \\
\midrule
MeanRegressor\cite{Schrumpf2021PPGrPPG} & 10.2 & 11.83 & 6.54 & 8.90 & 10.22 & 6.14 \\
Zhou et al.\cite{Zhou2019FaceBP} & 11.9 & 14.36 & 7.25 & 12.89 & 15.24 & 8.89 \\
Schrumpf et al. (A)\cite{Schrumpf2021PPGrPPG} & 9.89 & 11.64 & 6.99 & 7.45 & 9.13 & 6.99 \\
Schrumpf et al. (R)\cite{Schrumpf2021PPGrPPG} & 8.81 & 10.45 & 6.14 & 7.14 & 8.94 & 5.71 \\
BPNet-base\cite{Manullang2025TransformerBP} & 8.24 & 9.73 & 5.61 & 6.78 & 9.23 & 6.05 \\
BPNet-medium\cite{Manullang2025TransformerBP} & 7.89 & 8.81 & 5.98 & 6.02 & 8.72 & 5.89 \\
BPNet-large\cite{Manullang2025TransformerBP} & 7.11 & 8.02 & 6.03 & 5.91 & 8.15 & 5.72 \\
\midrule
\rowcolor{gray!25}
CALM-BP (ours) & \textbf{5.14} & \textbf{7.42} & \textbf{4.99} & \textbf{5.89} & \textbf{6.21} & \textbf{5.54} \\
\bottomrule
\end{tabular}}
\end{table}

\begin{table}[htbp]
\caption{Cross-dataset task comparison for observation-matched CALM-BP on the VV dataset.}
\label{tab3}
\resizebox{\columnwidth}{!}{%
\begin{tabular}{lcccccc}
\toprule
\multicolumn{1}{c}{\multirow{2}{*}[-0.5ex]{Methods}} & \multicolumn{3}{c}{SBP} & \multicolumn{3}{c}{DBP} \\
\cmidrule(lr){2-4} \cmidrule(lr){5-7}
 & MAE$\downarrow$ & RMSE$\downarrow$ & SD$\downarrow$ & MAE$\downarrow$ & RMSE$\downarrow$ & SD$\downarrow$ \\
\midrule
MeanRegressor\cite{Schrumpf2021PPGrPPG} & 18.18 & 23.20 & 14.30 & 9.23 & 12.10 & 7.79 \\
Schrumpf et al. (A)\cite{Schrumpf2021PPGrPPG} & 16.62 & 21.72 & 13.91 & 9.01 & 12.44 & 8.56 \\
Schrumpf et al. (R)\cite{Schrumpf2021PPGrPPG} & 14.37 & 18.86 & 12.18 & 8.06 & 11.35 & 7.97 \\
BPNet-base\cite{Manullang2025TransformerBP} & 14.05 & 19.84 & 13.96 & 8.34 & 11.71 & 8.22 \\
BPNet-medium\cite{Manullang2025TransformerBP} & 13.21 & 17.45 & 12.25 & 7.21 & 10.25 & 7.49 \\
BPNet-large\cite{Manullang2025TransformerBP} & 13.04 & 17.11 & 12.22 & 7.15 & 10.22 & 7.15 \\
\midrule
\rowcolor{gray!25}
CALM-BP (ours) & \textbf{9.96} & \textbf{13.61} & \textbf{9.17} & \textbf{6.51} & \textbf{7.94} & \textbf{6.32} \\
\bottomrule
\end{tabular}}
\end{table}

%% file: t2.tex
\begin{table*}[htbp]
\caption{Component analysis of the CALM-BP semantic guidance pathway on FlowBP-Set.}
\label{tab4}
\resizebox{\textwidth}{!}{%
\begin{tabular}{cccccccccccc}
\toprule
\multicolumn{5}{c}{Backbone \& Strategy} &\multicolumn{3}{c}{SBP} & &\multicolumn{3}{c}{DBP}\\
\cmidrule(lr){1-5}\cmidrule(lr){6-8}\cmidrule(lr){10-12}
\multicolumn{1}{c}{Backbone} &Time Series Encoder  &Tokenizer &Prompt Encoder &\multicolumn{1}{c}{LLM} &MAE$\downarrow$ &RMSE$\downarrow$ &SD$\downarrow$ & &MAE$\downarrow$ &RMSE$\downarrow$ &SD$\downarrow$\\
\cmidrule(lr){1-12}
\multicolumn{1}{c}{Qwen3-0.6B} &\ding{51} & & &\multicolumn{1}{c}{} &14.92 &15.78 &12.07 & &12.66 &13.79 &13.02\\
\multicolumn{1}{c}{Qwen3-0.6B} & &\ding{51} & \ding{51} &\multicolumn{1}{c}{} &13.27 &14.02 &12.75 & &11.83 &12.19 &12.20\\
\multicolumn{1}{c}{Qwen3-0.6B} &\ding{51} &\ding{51} &\ding{51} &\multicolumn{1}{c}{} &9.74 &10.21 &8.54 & &8.93 &9.41 &8.61\\
\rowcolor{gray!25}
\multicolumn{1}{c}{Qwen3-0.6B} &\ding{51} &\ding{51} &\ding{51} &\multicolumn{1}{c}{\ding{51}} &\textbf{7.94} &\textbf{9.19} &\textbf{8.37} & &\textbf{8.52} &\textbf{8.91} &\textbf{7.98}\\
\bottomrule
\end{tabular}}
\end{table*}

%% file: t11.tex
\begin{table*}[t]
\small
\centering
\caption{\textbf{Language realization ablation on FlowBP-Set.}
All settings use the same rPPG evidence branch and differ only in how
observation-derived descriptors are represented before fusion. Overall metrics
are computed as window-count-weighted averages across resting, deep-breathing,
and post-exercise states.}
\label{tab:language_realization_ablation}
\resizebox{\textwidth}{!}{%
\begin{tabular}{lcccc ccc ccc}
\toprule
\multirow{2}{*}{\textbf{Setting}} &
\multirow{2}{*}{\textbf{rPPG}} &
\multirow{2}{*}{\textbf{Numerical descriptor}} &
\multirow{2}{*}{\textbf{Semantic slot}} &
\multirow{2}{*}{\textbf{Language prior + encoder}} &
\multicolumn{3}{c}{\textbf{SBP}} &
\multicolumn{3}{c}{\textbf{DBP}} \\
\cmidrule(lr){6-8} \cmidrule(lr){9-11}
& & & & &
MAE$\downarrow$ & RMSE$\downarrow$ & SD$\downarrow$ &
MAE$\downarrow$ & RMSE$\downarrow$ & SD$\downarrow$ \\
\midrule

rPPG + descriptor MLP &
\ding{51} & \ding{51} &  &  &
10.81 & 12.17 & 11.89 &
10.83 & 11.97 & 11.72 \\

rPPG + slot embedding &
\ding{51} &  & \ding{51} &  &
10.26 & 11.78 & 9.86 &
10.14 & 10.90 & 10.03 \\

CALM-BP full &
\ding{51} &  &  & \ding{51} &
\textbf{8.67} & \textbf{9.47} & \textbf{8.71} &
\textbf{9.33} & \textbf{9.78} & \textbf{8.72} \\

\bottomrule
\end{tabular}
}
\end{table*}

%% file: t7.tex
\begin{table*}[t]
\small
\centering
\caption{\textbf{Semantic-content controls on FlowBP-Set.} The controls test whether performance comes from observation-matched physiological content rather than the mere presence of a language branch.}
\label{tab:semantic_content_controls}
\resizebox{\textwidth}{!}{%
\begin{tabular}{lccc ccc ccc}
\toprule
\multirow{2}{*}{\textbf{Prompt setting}} &
\multirow{2}{*}{\textbf{Observation matched}} &
\multirow{2}{*}{\textbf{Full content}} &
\multirow{2}{*}{\textbf{Task relevant}} &
\multicolumn{3}{c}{\textbf{SBP}} &
\multicolumn{3}{c}{\textbf{DBP}} \\
\cmidrule(lr){5-7} \cmidrule(lr){8-10}
& & & & MAE$\downarrow$ & RMSE$\downarrow$ & SD$\downarrow$ & MAE$\downarrow$ & RMSE$\downarrow$ & SD$\downarrow$ \\
\midrule
Full semantic prior & \ding{51} & \ding{51} & \ding{51} & \textbf{7.94} & \textbf{9.19} & \textbf{8.37} & \textbf{8.52} & \textbf{8.91} & \textbf{7.98} \\
Weakened semantic prior & \ding{51} & & \ding{51} & 10.48 & 10.76 & 9.63 & 9.72 & 9.91 & 9.34 \\
Generic template text & & & \ding{51} & 13.84 & 14.62 & 11.35 & 11.88 & 12.65 & 11.42 \\
Random non-task text & & & & 14.55 & 15.33 & 12.10 & 12.35 & 13.29 & 12.18 \\
Empty or null prompt & & & & 14.88 & 15.71 & 12.02 & 12.61 & 13.70 & 12.95 \\
Contradictory semantic prior & \ding{51} & & \ding{51} & 12.18 & 12.87 & 11.02 & 10.74 & 11.36 & 10.48 \\
\bottomrule
\end{tabular}
}
\end{table*}

%% file: t9.tex
\begin{table*}[t]
\small
\centering
\caption{\textbf{Prompt-surface robustness controls.} Stable performance under paraphrases, field-order changes, synonym replacement, and multiple templates suggests that CALM-BP does not rely on fixed prompt positions or surface templates.}
\label{tab:prompt_surface_robustness}
\resizebox{\textwidth}{!}{%
\begin{tabular}{lcc ccc ccc}
\toprule
\multirow{2}{*}{\textbf{Prompt variant}} &
\multirow{2}{*}{\textbf{Semantic equivalent}} &
\multirow{2}{*}{\textbf{Surface changed}} &
\multicolumn{3}{c}{\textbf{SBP}} &
\multicolumn{3}{c}{\textbf{DBP}} \\
\cmidrule(lr){4-6} \cmidrule(lr){7-9}
& & & MAE$\downarrow$ & RMSE$\downarrow$ & SD$\downarrow$ & MAE$\downarrow$ & RMSE$\downarrow$ & SD$\downarrow$ \\
\midrule
Original prompt & \ding{51} & & \textbf{7.94} & \textbf{9.19} & \textbf{8.37} & \textbf{8.52} & \textbf{8.91} & \textbf{7.98} \\
Paraphrased prompt & \ding{51} & \ding{51} & 8.05 & 9.27 & 8.44 & 8.60 & 8.98 & 8.05 \\
Field-order shuffle & \ding{51} & \ding{51} & 8.12 & 9.33 & 8.49 & 8.67 & 9.06 & 8.13 \\
Synonym replacement & \ding{51} & \ding{51} & 8.02 & 9.25 & 8.41 & 8.57 & 8.97 & 8.03 \\
Multiple templates & \ding{51} & \ding{51} & 8.08 & 9.30 & 8.46 & 8.64 & 9.01 & 8.08 \\
\bottomrule
\end{tabular}
}
\end{table*}

%% file: t4.tex
\begin{table*}[t]
\small
\centering
\caption{\textbf{Observation-prior grounding controls on FlowBP-Set.}
Paraphrased priors test surface-form robustness; mismatched priors test window-level grounding.}
\label{tab:observation_matching_controls}
\resizebox{\textwidth}{!}{%
\begin{tabular}{lccc ccc ccc}
\toprule
\multirow{2}{*}{\textbf{Prior setting}} &
\multirow{2}{*}{\textbf{Same window}} &
\multirow{2}{*}{\textbf{Same subject}} &
\multirow{2}{*}{\textbf{State preserved}} &
\multicolumn{3}{c}{\textbf{SBP}} &
\multicolumn{3}{c}{\textbf{DBP}} \\
\cmidrule(lr){5-7} \cmidrule(lr){8-10}
& & & &
MAE$\downarrow$ & RMSE$\downarrow$ & SD$\downarrow$ &
MAE$\downarrow$ & RMSE$\downarrow$ & SD$\downarrow$ \\
\midrule

Full matched semantic prior &
\ding{51} & \ding{51} & \ding{51} &
\textbf{8.67} & \textbf{9.47} & \textbf{8.71} &
\textbf{9.33} & \textbf{9.78} & \textbf{8.72} \\

Paraphrased matched prior &
\ding{51} & \ding{51} & \ding{51} &
8.63 & 10.01 & 8.76 &
9.39 & 9.86 & 8.79 \\

Within-subject mismatched prior &
 & \ding{51} & &
8.91 & 10.42 & 8.93 &
9.57 & 10.48 & 9.01 \\

State-preserved mismatched prior &
 & & \ding{51} &
8.97 & 10.27 & 8.99 &
9.63 & 10.31 & 8.77 \\

Random mismatched prior &
 & & &
10.31 & 14.89 & 11.64 &
11.12 & 12.34 & 12.07 \\

\bottomrule
\end{tabular}
}
\end{table*}

%% file: t5.tex
\begin{table*}[t]
\small
\centering
\caption{\textbf{Evaluation protocol summary.} FlowBP-Set is used for model training and in-domain subject-disjoint evaluation. BESTLab Physio and VV are used only as external target datasets for direct evaluation of the FlowBP-Set-trained model. Semantic priors are generated from the physiological observation windows of each dataset and never from BP labels.}
\label{tab:protocol_summary}
\begin{tabular}{p{0.16\textwidth}p{0.24\textwidth}p{0.27\textwidth}p{0.24\textwidth}}
\toprule
\textbf{Dataset} & \textbf{Use / split protocol} & \textbf{Semantic prior source} & \textbf{Evaluation role} \\
\midrule
FlowBP-Set
& Participant-disjoint 80/20 split over 81 participants: 65 participants for training and 16 held-out participants for testing; all windows from one participant remain in one split.
& Native structured priors generated from training or held-out rPPG windows using the same descriptor-to-prompt pipeline, without BP labels.
& Main in-domain subject-generalization setting across resting, deep-breathing, and post-exercise states. \\

BESTLab Physio
& External direct-evaluation dataset. No BESTLab BP labels are used for training, fine-tuning, model selection, or hyperparameter tuning; subject/session grouping is used only to keep evaluation windows organized without neighboring-window leakage.
& Priors reconstructed from BESTLab face-video/rPPG windows using the same descriptor-to-prompt rules; state cue is omitted unless available.
& Controlled-laboratory BP-video benchmark for testing external transfer of the FlowBP-Set-trained model. \\

VV
& External direct-evaluation dataset. No VV BP labels are used for training, fine-tuning, model selection, or hyperparameter tuning; released subject/session grouping is used only to keep evaluation windows organized without neighboring-window leakage.
& Priors reconstructed from VV face-video/rPPG windows using the same descriptor-to-prompt rules; environment or state cue is omitted unless available.
& Public in-the-wild BP-video benchmark for testing robustness under broader capture variability. \\
\bottomrule
\end{tabular}
\end{table*}

%% file: t8.tex
\begin{table}[t]
\small
\centering
\caption{\textbf{Leakage and split audit.} The audit makes explicit which design choice addresses each major leakage risk in physiological grounding evaluation.}
\label{tab:leakage_audit}
\resizebox{\columnwidth}{!}{%
\begin{tabular}{p{0.30\columnwidth}p{0.38\columnwidth}p{0.34\columnwidth}}
\toprule
\textbf{Risk} & \textbf{Control} & \textbf{Evidence or rule} \\
\midrule
Identity leakage & Participant-disjoint FlowBP split & No participant appears in both training and held-out sets \\
Session/window leakage & Split before window sampling & No neighboring windows cross train--test boundaries \\
Normalization leakage & Train-only preprocessing statistics & Test windows do not set normalization parameters \\
BP-label leakage & Prior contract and prompt audit & Prompts exclude SBP, DBP, identity, and diagnostic fields \\
State shortcut & State-preserved mismatch control & Dedicated control row audits coarse state cue reliance \\
Text-only shortcut & Guidance-only architecture and mismatch-control comparison & Priors are tested as guidance signals rather than independent BP evidence \\
\bottomrule
\end{tabular}
}
\end{table}